\documentclass[pre,twocolumn,amsmath,amssymb,floatfix]{revtex4}

\usepackage{graphicx}
\usepackage{dcolumn} 
\usepackage{bm}      
\usepackage{color}

\begin{document}

\title{The decay of Batchelor and Saffman rotating turbulence}
\author{Tomas Teitelbaum$^{1}$ and Pablo D.~Mininni$^{1,2}$}
\affiliation{$^1$ Departamento de F\'\i sica, Facultad de Ciencias 
Exactas y Naturales, Universidad de Buenos Aires, \& IFIBA, CONICET, 
Ciudad Universitaria, 1428 Buenos Aires, Argentina. \\
$^2$ NCAR, P.O. Box 3000, Boulder, Colorado 80307-3000, U.S.A.}
\date{\today}

\begin{abstract}
The decay rate of isotropic and homogeneous turbulence is known to be 
affected by the large-scale spectrum of the initial perturbations, 
associated with at least two cannonical self-preserving solutions of the 
von K\'arm\'an-Howarth equation: the so-called Batchelor and Saffman 
spectra. The effect of long-range correlations in the decay of
anisotropic flows is less clear, and recently it has been proposed
that the decay rate of rotating turbulence may be independent of the 
large-scale spectrum of the initial perturbations. We analyze numerical 
simulations of freely decaying rotating turbulence with initial energy 
spectra $\sim k^4$ (Batchelor turbulence) and $\sim k^2$ (Saffman 
turbulence) and show that, while a self-similar decay cannot be 
identified for the total energy, the decay is indeed affected by 
long-range correlations. The decay of two-dimensional and 
three-dimensional modes follows distinct power laws in each case,
which are consistent with predictions derived from the anisotropic 
von K\'arm\'an-Howarth equation, and with conservation of anisotropic 
integral quantities by the flow evolution.
\end{abstract}
\maketitle

\section{Introduction}

Turbulent flows subject to background rotation are an important problem 
in fluid mechanics as several systems are affected by rotation. Rotation 
affects the Earth atmosphere and the oceans at large scales, it is crucial 
in many engineering flows such as in turbomachinery, and is also 
important in many astrophysical systems such as stellar convective 
regions and gaseous planets atmospheres. As many applications arise in 
nature and technology, it becomes important to understand 
detailed properties of these flows.

In particular, a long-lasting problem in fluid mechanics is that of the 
decay of turbulent fluctuations. The rate of energy decay in turbulent 
flows is known to be sensitive to initial conditions, and no single 
universal solution to the decay appears to exist to which all flows 
converge asymptotically for infinite Reynolds number \cite{George1992}. 
Considering isotropic and homogeneous turbulence at high Reynolds number, 
there are two well known canonical cases as far as the energy spectrum 
at large scales is concerned. Provided the cubic velocity correlation 
tensor $\left<u_i({\bf x}) u_j({\bf x}) u_k({\bf x}+{\bf r})\right>$
decays sufficiently fast for large $r$, these cases correspond to 
the so-called Saffman spectrum for which $E(k \rightarrow 0) \sim Lk^2$, 
and to the so-called Batchelor spectrum where 
$E(k \rightarrow 0) \sim Ik^4$. Which of these spectra is observed 
depends on the initial conditions, and other non-canonical cases may arise 
if the above condition on the decay of the correlation tensor does not
hold (see, e.g., \cite{Krogstad2010}). The quantities $L$ and $I$ are 
approximately conserved integrals for fully developed, freely decaying 
turbulence in each case ($I$ is often called the Loytsiansky
integral), and have been used to determine the decay rate of the total 
energy (see \cite{Batchelor1956,Saffman1967,Saffman1967b,Ishida2006} 
for examples). The quasi-conservation in each case is associated with 
self-preserving solutions of the von K\'arm\'an-Howarth equations, 
and can also be interpreted in terms of conservation of linear and
angular momentum. While constancy of $L$ is a consequence of linear 
momentum conservation \cite{Saffman1967}, constancy of $I$ is a 
consequence of the conservation of angular momentum 
\cite{Landau1959,Davidson2009}.

The presence of background rotation breaks down isotropy as a preferred 
direction arises along the axis of rotation (for detailed studies 
of rotating turbulence, see \cite{Cambon1989,Cambon1997,Bellet2006}). 
Flows subjected to rotation develop anisotropies which have been 
shown to impact dramatically on the decay of energy, and there are 
different and sometimes conflicting results in the literature
regarding the decay laws followed by the energy in rotating flows. 
Moreover, theoretical arguments based on the isotropic von 
K\'arm\'an-Howarth equations are not valid anymore, and must be 
extended to include the effect of rotation. The integrals $I$ and $L$ 
become tensors, and for axisymmetric flows some of the diagonal 
components of these tensors can be expected to replace $I$ and $L$ 
as approximately conserved quantities. Extensions of these arguments 
to anisotropic cases such as conducting flows with an imposed 
magnetic field, rotating, or stratified flows, have been recently
derived in Refs.~\cite{Davidson2009,Davidson2010,Okamoto2010}.

The development of anisotropies in rotating flows have been reported 
both in experiments and simulations. Experiments of freely decaying 
rotating turbulence show a reduction of the energy transfer, evidenced 
by a decrease in the energy decay rate \cite{Bokhoven2009}. An 
anisotropic energy flux with a trend towards 
quasi-two-dimensionalization has been recently reported 
\cite{Lamriben2011}, and is also evidenced, e.g., by an increase in the 
correlation lengths along the axis of rotation 
\cite{Baroud2003,Staplehurst2008}. Regarding the energy decay, 
experimental results for non-rotating grid turbulence with integral 
scales smaller than the size of the vessel \cite{Morize2006} show an 
energy decay $E \sim t^{-1.1}$ compatible with the theoretical result 
$E \sim t^{-6/5}$ expected for large scale initial conditions with 
Saffman spectrum $\sim k^2$. In \cite{Moisy2011}, a transition 
from a $E \sim t^{-6/5}$ decay to a decay law closer to 
$\sim t^{-3/5}$ is reported once rotation starts affecting the flow 
dynamics, together with a cyclone-anticyclone asymmetry evidenced by 
positive skewness which grows with rotation rate. The decay for bounded 
flows, in which initial integral scales are close to the size of the 
vessel, seems to change from $E \sim t^{-2}$ to $\sim t^{-1}$ when rotation 
is present \cite{Morize2006}.

Numerical simulations and models also show a slow down of the energy 
decay for many different initial conditions 
\cite{Jacquin1990,Yang2004,Bellet2006,Muller2007,Bokhoven2008,Teitelbaum2009}, 
together with the trend towards two-dimensionalization 
\cite{Muller2007,Bourouiba2007}, and the higher cyclonic-over-anticyclonic 
activity \cite{Teitelbaum2010,POF}. The first theoretical 
study of the decay of rotating turbulence was reported in 
\cite{Cambon1994}, and considered the decrease of the energy transfer
in the presence of rotation to explain the observed slow down in the
decay, and to predict decay rates for the energy. The exponents
reported were consistent with those obtained in simulations in 
\cite{Jacquin1990}, and in the experiments in \cite{Morize2006}. Also,
a very close decay law $E \sim t^{-0.8}$ was found in
\cite{Bellet2006} for a wide range of Rossby numbers (from very small 
to moderate values), when solving equations for inertial wave turbulence
and for an Eddy Damped Quasi-Normal Markovian model (EDQNM3).

Later, in \cite{Bourouiba2007} a distinction was introduced between 
the energy contained in modes with $k_z=0$ ($E_{2D}$, corresponding 
to two-dimensional, ``slow'' or ``vortical'' modes) and modes with 
$k_z \neq 0$ ($E_{3D}$, corresponding to three-dimensional, ``fast'', 
or ``wave'' modes). The authors carried a comprehensive study of 
the transfer of energy from 3D to 2D modes, for a wide range of Rossby 
numbers, and found a non-monotonic behavior for the decay of energy 
for large, intermediate, and small Rossby numbers. They focused on the 
intermediate Rossby range, where the maximal energy transfer between 3D  
and 2D modes occurs (see also \cite{Bourouiba2011}). In this range,
they reported an initial growth of $E_{2D}$ as soon as the decay
begins, that eventually results in a cross-over with the 
monotonously decaying $E_{3D}$. Reference \cite{Thiele2009} 
also reported different behavior in the time evolution of $E_{2D}$ and 
$E_{3D}$, but considered only the decay of the total energy 
$E = E_{2D} + E_{3D}$ proposing a decay that depends on the amount of 
background rotation $E \sim t^{-\gamma(\Omega)}$. The dependence of 
$\gamma$ on the rotation rate was derived using phenomenological 
arguments resulting $\gamma\sim \Omega^{-1}$.

This variety of solutions naturally led to parametric studies of the
initial condition space in numerical simulations. For non-helical 
rotating flows with initial integral scale close to the size of the 
domain \cite{Teitelbaum2009}, a decay $E \sim t^{-1}$ was found in 
simulations, in agreement with the experimental results \cite{Morize2006}. 
For flows with initial integral scale sufficiently smaller than the 
domain size, the case of initial conditions with energy spectra 
$\sim k^4$ was studied in detail in Refs.~\cite{Teitelbaum2010,POF}, 
where different decay rates were reported for the two-dimensional and 
three-dimensional modes. While $E_{3D}$ was found to decay as in the 
case of isotropic turbulence, a slower decay was found for $E_{2D}$ in 
the intermediate-Rossby range as defined in \cite{Bourouiba2007}. The 
decay of two-dimensional modes can be correctly explained if only a few 
components of the Loytsiansky tensor are considered (here referred in 
short as $I_{\perp}$ and $I_{\parallel}$ based on the 
symmetries of the flow), and it was shown that $I_{\perp}$ remains 
approximately constant during the decay \cite{POF}. Finally, it was also 
found that the decay is influenced by the amount of total helicity 
present in the initial conditions, and by the initial degree of 
anisotropy in the flow.

Recently, the anisotropic von K\'arm\'an-Howarth equation was used to 
consider the decay of a rotating turbulent flow with initial conditions 
following a Saffman spectrum \cite{Davidson2010}. The approximately 
conserved integral $L$ in the isotropic and homogeneous case, was 
generalized in the axisymmetric case to two integrals 
$L_{\perp} \propto L_{xx} = L_{yy}$ and 
$L_{\parallel} = L_{zz}$. Assuming that both $L_{\perp}$ and $L_{\parallel}$ 
are simultaneously conserved (or $I_\perp$ and $I_{||}$ for 
the case of Batchelor initial 
spectrum), and using the empirical result that characteristic lengthscales 
of the flow in the direction parallel to the rotation axis grow linearly 
with time \cite{Davidson2010}, the author concludes that regardless of 
the initial spectrum being $\sim k^2$ or $\sim k^4$, both cases should 
decay with the same $E \sim t^{-1}$ law, thus implying that it is not 
possible to set apart $\sim k^2$ from $\sim k^4$ turbulence by 
measuring the energy decay exponent alone.

In this paper we analyze whether the decay of rotating flows is affected 
by long-range correlations, studying the energy decay rate for flows with 
$\sim k^2$ and $\sim k^4$ initial conditions. We first present a brief 
theoretical discussion of conserved integral quantities for rotating 
turbulence using a von K\'arm\'an-Howarth equation which includes a 
Coriolis term due to rotation. We extend the previously derived results 
for Saffman turbulence \cite{Davidson2010}, considering Batchelor 
spectrum. The results corroborate the phenomenological 
arguments used in \cite{POF} to explain the decay laws found numerically 
for energy spectra $k^4$. In the second part of the paper we use 
numerical simulations to analyze the decay of rotating Saffman and 
Batchelor turbulence in periodic domains. We focus on the regime of 
intermediate Rossby numbers (initial $Ro \approx 0.1$ \cite{Bourouiba2007}), 
and therefore we will not attempt a study of the dependence of the decay 
with the rotation rate. To the best of our knowledge, the free decay of 
energy for $\sim k^2$ initial conditions has not yet been numerically 
studied for rotating turbulence, and the resulting decay laws seem 
to be relevant for experiments. We show that the decay of energy in 
two-dimensional and three-dimensional modes follows distinct power 
laws for the cases of Batchelor and Saffman turbulence. The decay 
laws are consistent with the predictions derived from the anisotropic 
von K\'arm\'an-Howarth equation, and with the approximate constancy of 
either $L_\perp$ or $I_\perp$ as the flow decays.

\section{\label{sec:Karman}von K\'arm\'an-Howarth equations and 
self-similar decay}

The dynamics of an incompressible fluid subjected to background 
rotation is described by the Navier-Stokes equation with the addition of 
the Coriolis acceleration,
\begin{equation}
\partial_t {\bf u} + \mbox{\boldmath $\omega$} \times
    {\bf u} + 2 \mbox{\boldmath $\Omega$} \times {\bf u}  =
    - \nabla {\cal P} + \nu \nabla^2 {\bf u} ,
\label{eq:momentum}
\end{equation}
together with the incompressibility condition,
\begin{equation}
\nabla \cdot {\bf u} = 0.
\label{eq:incompressible}
\end{equation}
Here, ${\bf u}$ is the velocity field, $\mbox{\boldmath $\omega$} = 
\nabla \times {\bf u}$ is the vorticity, the centrifugal acceleration 
is absorbed in the total pressure per unit of mass ${\cal P}$, and 
$\nu$ is the kinematic viscosity. We assume uniform density and the 
rotation axis in the $z$ direction so $\mbox{\boldmath $\Omega$} = 
\Omega \hat{z}$, with $\Omega$ the rotation frequency.

We briefly present the derivation of the von K\'arm\'an-Howarth equation 
including the Coriolis term in order to find invariant quantities during 
the self-similar energy decay in rotating flows. To this end, we write 
the Navier-Stokes equation in a rotating frame in index notation, and 
evaluated at two points ${\bf x}$ and ${\bf x}'={\bf x}+{\bf r}$,
\begin{equation}
\frac{\partial u_i}{\partial t}=-\frac{\partial}{\partial x_k}(u_iu_k)-
    \frac{\partial p}{\partial x_i} + \nu \nabla^2u_i - 
    2\epsilon_{imn}\Omega_mu_n ,
\label{eq:NSROT1}
\end{equation}
and
\begin{equation}
\frac{\partial u'_j}{\partial t}=-\frac{\partial}{\partial x'_k}(u'_ju'_k)-
    \frac{\partial p'}{\partial x'_j} + \nu \nabla'^2u'_j - 
    2\epsilon_{jmn}\Omega_mu'_n ,
\label{eq:NSROT2}
\end{equation}
where ${\bf u}'={\bf u}({\bf x}')$, $\nabla'^2$ denotes 
Laplacian with respect to the ${\bf x}'$ coordinate, and $\epsilon$ is
the Levi-Civita symbol. Multiplying Eq.~(\ref{eq:NSROT1}) by $u'_j$, 
Eq.~(\ref{eq:NSROT2}) by $u_i$, summing and averaging, we get the 
equation for the time evolution of the two-point velocity 
correlation tensor,
\begin{eqnarray}
&&\frac{\partial}{\partial t} \left<u_iu'_j\right> = 
    -\left( \frac{\partial}{\partial x_k}
    \left< u'_ju_iu_k \right> +\frac{\partial}{\partial x'_k}
    \left<u_iu'_ju'_k \right> \right)
    \nonumber \\
{} && - \left( \frac{\partial}{\partial x_i} \left< u'_jp \right> + 
    \frac{\partial}{\partial x'_j} \left< u_ip' \right> \right) + 
    \nu \left( \nabla^2 \left< u'_ju_i \right> + \right.
    \nonumber \\
{} && + \left. \nabla'^2 \left< u_iu'_j \right> \right) 
    - 2\Omega_m \left<\epsilon_{imn}u_nu'_j+\epsilon_{jmn}u'_nu_i\right>.
\label{eq:KH}
\end{eqnarray}

If $\Omega = 0$, and assuming the cubic velocity correlation tensor 
$\left<u_iu_ju'_k\right>$ and the pressure-velocity correlation 
$\left<u_i p' \right>$ decay fast enough with $r$ when 
$r \rightarrow \infty$, two possible integral conserved quantities can 
be obtained from Eq.~(\ref{eq:KH}). When the terms on the r.h.s.~of the 
equation go to zero as $\mathcal{O}(r^{-2})$, then it follows from the 
time derivative on the l.h.s.~that 
\begin{equation}
L={\int}\left<{\bf u\cdot u}'\right>d{\bf r}
\label{eq:L}
\end{equation}
is conserved. This is know as the Saffman integral \cite{Saffman1967}.
Multiplying 
Eq.~(\ref{eq:KH}) by $r^2$, another possible invariant is obtained
when the terms on the l.h.s.~are $\mathcal{O}(r^{-4})$, namely
\begin{equation}
I=-{\int}r^2\left<{\bf u\cdot u}'\right>d{\bf r},
\end{equation}
which is known as the Loitsyanski integral and whether it actually 
remains constant or not in decaying isotropic and homogeneous turbulence 
has been a matter of debate \cite{Davidson2004}. As the energy 
spectrum for small values of $k$ can be expanded as
\begin{equation}
E(k \rightarrow 0) \sim Lk^2+Ik^4+... \, ,
\label{eq:Elowk}
\end{equation}
the integrals $L$ or $I$ are then expected to be quasi-invariants during 
the decay for initial large-scale energy spectra of the form $\sim k^2$ 
(Saffman spectrum) \cite{Saffman1967,Saffman1967b} or $\sim k^4$ 
(Batchelor spectrum) \cite{Batchelor1956} respectively. Using these 
quasi-conserved integrals, the energy balance equation 
$dE/dt \sim -E^{3/2}/l$ can be closed (as either $l^3u^2$ or $l^5u^2$ are 
constant, with $l$ the energy-containing scale), and different decay laws 
arise for each scaling of the initial spectra,
\begin{equation}
E(t) \sim \left\{ \begin{array}{ll}
 t^{-6/5} & \textrm{if $E(k,t=0) \sim k^2$}\\
 t^{-10/7} & \textrm{if $E(k,t=0) \sim k^4$}
\end{array} \right.
\label{eq:teodecay}
\end{equation}
These decay laws has been observed in numerical as well as in experimental 
studies (see, e.g., 
\cite{Kolmogorov1941,Ishida2006,Saffman1967,Saffman1967b,Morize2006,POF,Moisy2011}) 

These isotropic integrals have been also assumed to remain constant during 
the decay of rotating turbulence with $\sim k^2$ and $\sim k^4$ large-scale 
energy spectra, to predict energy decay laws \cite{Cambon1994}. In the 
presence of rotation, the energy per unit of time (flux) transferred 
towards smaller scales is reduced as $\Omega$ is increased. A 
phenomenological expression consistent with the observations is to assume 
that the flux is reduced by the ratio of the turnover time 
$\tau \sim l/E^{1/2}$ to the wave time $(2 \Omega)^{-1}$ 
\cite{Iroshnikov1963,Kraichnan1965,Muller2007,Mininni2009}, 
resulting in \cite{Cambon1994}
\begin{equation}
\frac{dE}{dt} \sim -\frac{E^2}{\Omega l^2}
\label{eq:balancerot}
\end{equation}
Further assuming $I$ or $L$ are conserved results in the following 
decays,

\begin{equation}
E(t) \sim \left\{ \begin{array}{ll}
 t^{-3/5} & \textrm{if $E(k,t=0) \sim k^2$}\\
 t^{-5/7} & \textrm{if $E(k,t=0) \sim k^4$}
\end{array} \right.
\label{eq:teodecay2}
\end{equation}
Although these power laws are close to the behavior found in
experiments and simulations, as mentioned in the introduction a better
agreement with observations can be obtained when the arguments are 
extended to consider the effect of anisotropy in the flow.

In the anisotropic case, and assuming absence of long-range correlations, 
only the final term in Eq.~(\ref{eq:KH}) can contribute to the rate of change 
of integrals of the type
\begin{equation}
I_{ijmn}={\int}r_mr_n\left<u_iu'_j\right>d{\bf r},
\label{Iijmn}
\end{equation}
for the Batchelor case, or
\begin{equation}
L_{ij}={\int}\left<u_iu'_j\right>d{\bf r}
\label{Lij}
\end{equation}
for Saffman turbulence. It is important to point out here that 
it is not trivial that in this case long-range correlations should
vanish as in the isotropic case (specially for the pressure velocity 
correlations, see \cite{Bellet2006} where these terms are computed 
using the Fourier transform of Eq.~(\ref{eq:KH}) and the Poisson 
equation for the pressure fluctuations). Thus, the validity of the 
assumptions can only be verified {\it a posteriori} from the results 
obtained from the numerical simulations.

Assuming long-range correlations vanish sufficiently fast at 
large-scales, the problem is that of finding what components of 
the tensors in Eqs.~(\ref{Iijmn}) and (\ref{Lij}) are still conserved 
in the presence of rotation. We derive here the Batchelor case 
($I_{ijmn}$) under the same assumptions for the decay of velocity 
correlations with $r$ as in the isotropic case, multiplying 
Eq.~(\ref{eq:KH}) by $r_m r_n$ and integrating to obtain
\begin{eqnarray}
&& \frac{\partial}{\partial t}{\int}r_mr_n\left<u_iu'_j\right>d{\bf r} = 
    \nonumber \\ 
{} && -2 \Omega_l {\int} r_mr_n \left( \epsilon_{ilk}\left< u_ku'_j\right> + 
    \epsilon_{jlk}\left<u'_ku_i\right> \right) d{\bf r} .
\end{eqnarray}
Let's assume $\Omega_l = \Omega_z \delta_{lz}$, so
\begin{eqnarray}
&& \frac{\partial}{\partial t}{\int}r_mr_n\left<u_iu'_j\right>d{\bf r} = 
    \nonumber \\
{} && -2 \Omega_z {\int} r_mr_n \left(\epsilon_{izk}\left<u_ku'_j\right> + 
    \epsilon _{jzk}\left<u'_{k}u_i\right> \right) d{\bf r} .
\end{eqnarray}
As the flow has axisymmetry, we are interested in the diagonal components 
of the tensor. For $j=i$,
\begin{equation}
\Omega_z\epsilon_{izk}\left<u_ku'_i\right>=\Omega_z\left<u_xu'_y-u_yu'_x\right>,
\end{equation}
\begin{equation}
\Omega_z\epsilon_{jzk}\left<u'_{k}u_j\right>=\Omega_z\left<u'_xu_y-u'_yu_x\right>,
\end{equation}
where all terms with $i=z$ or $j=z$ vanish. For $m=n$, introducing 
${\bf r_{\perp}} = (r_x,r_y,0)$ and ${\bf u}_{\perp}=(u_x,u_y,0)$ we get
\begin{eqnarray}
&& \frac{\partial}{\partial t}{\int}r_{\perp}^2\left<u_{\perp}u'_{\perp}\right>
    d{\bf r} = 
    \nonumber \\
{}&& -2{\int}r_{\perp}^2\Omega_z[\left<u_xu'_y-u_yu'_x\right>-
    \left<u_xu'_y-u_yu'_x\right>]d{\bf r},
\end{eqnarray}
so the terms on the r.h.s.~of the equation cancel and
\begin{equation}
\frac{\partial}{\partial t}{\int}r_{\perp}^2\left<
    {\bf u}_{\perp}{\bf u}'_{\perp}\right>d{\bf r} = 0,
\end{equation}
or equivalently,
\begin{equation}
I_{\perp} = {\int}r_{\perp}^2\left<{\bf u}_{\perp}{\bf u}'_{\perp}\right>d{\bf r} 
    = \textrm{constant} .
\label{eq:I2D}
\end{equation}

The case of Saffman turbulence was considered in \cite{Davidson2010}. The 
relevant components of the tensor $L_{ij}$ for axisymmetric flows are 
$L_{xx}=L_{yy}$ and $L_{zz}$. Similar arguments as the ones described 
above lead, for the $xx$ and $yy$-components of the von K\'arm\'an-Howarth 
equation, to the conservation of the integral
\begin{equation}
L_{\perp} = {\int}\left<{\bf u}_{\perp}{\bf u}'_{\perp}\right>d{\bf r} = 
    \textrm{constant} .
\label{eq:K}
\end{equation}

Note that both $I_{\perp}$ and $L_{\perp}$ resemble integrals that arise in the 
context of the decay of $2D$ turbulence \cite{Fox2009}. This is 
to be expected as the spectrum of an axisymmetric flow can be expanded, 
for the modes with $k_\parallel=k_z=0$, as
\begin{equation}
E(k_{\perp} \rightarrow 0,k_{\parallel}=0) \approx 
    L_{\perp}k_{\perp}+I_{\perp}k^3_{\perp}+...
\label{eq:expansion2d}
\end{equation}
which is (except for differences in the dimensions) also the expansion
of a 2D energy spectra. The resemblance can be expected as rotating
flows tend to become quasi-2D, concentrating most of the energy in 
the slow modes with $k_\parallel=0$.

For axisymmetric turbulence and a Batchelor initial spectrum $\sim k^4$ 
at large scales, Eq.~(\ref{eq:I2D}) leads to
\begin{equation}
I_{\perp} \sim l_{\perp}^4l_{\parallel}u_{\perp}^2 \approx\textrm{constant} ,
\label{eq:Iperp}
\end{equation}
where $l_\perp$ and $l_\parallel$ are characteristic (energy-containing) 
scales in the perpendicular and parallel direction respectively. Assuming 
2D and 3D modes are only weakly coupled in rotating turbulence, from 
Eq.~(\ref{eq:balancerot}) we can write an equation for the decay of the 
energy in 2D modes
\begin{equation}
\frac{dE_{2D}}{dt} \sim -\frac{E_{2D}^2}{\Omega I_{\perp}^2} ,
\label{eq:E2D}
\end{equation}
and from Eq.~(\ref{eq:Iperp})
\begin{equation}
\frac{dE_{2D}}{dt} \sim -\frac{E_{2D}^{5/2}l_{\parallel}^{1/2}}
    {I_{\perp}^{1/2}\Omega}.
\label{eq:balancerot2D}
\end{equation}
Assuming $E_{2D} \sim t^\gamma$ it follows that \cite{POF} 
\begin{equation}
E_{2D} \sim t^{-2/3}
\end{equation} 
if $l_\parallel$ remains constant (which is reasonable as 2D modes have 
no dependence on the direction parallel to rotation, and therefore 
$l_\parallel = l_0$ is the vertical size of the box).

For a $\sim k^2$ initial spectra,
\begin{equation}
L_{\perp} \sim l_{\perp}^2l_{\parallel}u_{\perp}^2 \approx \textrm{constant} ,
\end{equation}
and replacing in Eq.~(\ref{eq:E2D}) leads to 
\begin{equation}
\frac{dE_{2D}}{dt} \sim -\frac{E_{2D}^3l_{\parallel}}{L_{\perp}\Omega} ,
\label{eq:balancerot2D2}
\end{equation}
and a decay 
\begin{equation}
E_{2D} \sim t^{-1/2}.
\end{equation}

The results for the decay of energy 
in 2D modes in rotating turbulence can be summarized as follows,
\begin{equation}
E_{2D}(t) \sim \left\{ \begin{array}{ll}
 t^{-1/2} & \textrm{if $E(k,t=0) \sim k^2$}\\
 t^{-2/3} & \textrm{if $E(k,t=0) \sim k^4$.}
\end{array} \right.
\label{eq:teodecayrot}
\end{equation}

It is worth mentioning that the decay of rotating turbulence with 
$\sim k^2$ spectra is discussed in detail in \cite{Davidson2010}, where 
the constancy of $L_{\perp}$ is obtained from the K\'arm\'an-Howarth equation. 
In that case, the author also considers the von K\'arm\'an-Howarth 
equation for the $zz$-component of the two-point correlation tensor, and 
assuming cubic correlations decay fast enough also for that component, 
obtains constancy of $L_\parallel \sim u_{\parallel}^2l_{\perp}^2l_{\parallel}$ 
as well as of $L_{\perp}$. The constancy of these two quantities, together with 
the empirical result $l_{\parallel}=l_0(1+\kappa \Omega t)$ (where 
$\kappa$ is a constant of order one \cite{Staplehurst2008,Jacquin1990}) 
leads to a decay $E_{2D}(t)\sim t^{-1}$. A similar argument for Batchelor 
$\sim k^4$ turbulence leads to the same decay, which would result in an 
impossibility to distinguish between $\sim k^2$ and $\sim k^4$ turbulence 
by measuring the energy decay exponent alone.

Finally, it is interesting to point out that constancy of $I_{\perp}$ and 
$L_{\perp}$ can be respectively associated with conservation of the 
$z$-component of angular and linear momentum (see 
\cite{Davidson2009,Davidson2010}). 
In fact, the mean squared angular momentum $H_z$ averaged over a 
cylinder-shaped volume $V$ is
\begin{equation}
\left<H_z^2\right> = 2\pi V l_\parallel I_{\perp},
\end{equation}
while the linear momentum $P_z$ is
\begin{equation}
\left<L_z^2\right> = 2\pi V l_\parallel L_{\perp}.
\end{equation}

\section{\label{sec:NumSim}Numerical Simulations}

In the following we resort to numerical simulations to verify the validity 
of the assumptions and predictions discussed in the previous section. 
As a well resolved initial spectrum $\sim k^2$ or $\sim k^4$ is needed 
to observe constancy of $I$ or $L$ in simulations of isotropic and 
homogeneous turbulence (see, e.g., \cite{Ishida2006}), we use 
large-eddy simulations (LES) in periodic boxes to be able to have
initial energy containing wavenumber $k_0 \approx 40$ with large 
Reynolds numbers at a reasonable computing cost.

Before proceeding, we must caution the reader on how the 
simulations in periodic boxes should be interpreted. Flows in periodic
boxes are sometimes considered as describing a homogeneous flow in an 
infinitely periodic domain. In this work, we will interpret instead
a flow in a periodic box as artificially confined. Indeed, as was
shown in \cite{POF}, when the integral scale of the flow is close to the
domain size, the eddies cannot become larger and the energy decay laws
obtained are those of confined flows. The confinement is artificial in
the sense that the flows have no Ekman layers, as the domain has no
rigid walls. Moreover, the confinement also results in a discrete set of 
wavenumbers and selects a discrete set of inertial waves which are
normal modes of the domain, unlike homogeneous rotating flows in 
infinite domains which have continuous wavenumbers (see, e.g., 
\cite{Smith2005,Bellet2006}). Note that due to the discretization of 
wavenumbers in the former case, the existence and number of exact 
resonances depend on the domain size \cite{Smith2005} (as this is 
aggravated when boxes with large aspect ratio and lower resolution 
in the vertical direction are used, we restrict our study to boxes with
aspect ratio of unity).

While these are unavoidable and inherent properties of the 
simulations in periodic domains, the choice of a relatively large 
initial energy containing wavenumber $k_0 \approx 40$ 
ensures that as the equations are advanced in time the energy
containing scale of the flow will remain smaller than the box size,
and that a large fraction of wavenumbers will be available at 
large-scales for the system to evolve.

\subsection{Equations and model}

Equations (\ref{eq:momentum}) and (\ref{eq:incompressible}) are solved 
numerically using a dynamical subgrid-scale spectral LES model of 
rotating turbulence in which only the large scales are explicitly 
resolved. The statistical effect of unresolved scales on the scales 
larger than a cut-off wave number ($k_c=N/2$) are modeled with an 
eddy-viscosity and eddy-noise that are obtained after solving in each 
time step the eddy damped quasi-normal Markovian (EDQNM) equations for 
the spectrum of the unresolved scales. Details of the LES can 
be found in Ref.~\cite{Baerenzung2008}, and in 
Refs.~\cite{Baerenzung2010,Baerenzung2011} 
for its extension to the rotating case. A validation of 
the LES against direct numerical simulations (DNS) for the case of 
freely decaying rotating turbulence can be found in \cite{POF}.

The simulation domain is a 3D periodic box of length $2\pi$ with 
spatial resolution of $256^3$ grid points (resulting in $k_c=128$). A 
parallelized pseudo-spectral method without de-aliasing is used to 
solve spatial derivatives, and an explicit second-order Runge-Kutta 
method is used to evolve in time \cite{Gomez1}.

\subsection{Initial conditions}

In \cite{POF} it was observed that the energy decay is sensitive 
to the initial anisotropy of the flow (in particular, to the initial 
amount of energy in 2D modes). To enforce spectral and variance 
isotropy of the initial conditions, we generate at $t=0$ a random 
velocity field using the Craya-Herring decomposition 
\cite{Craya1958,Herring1974} which associates to each wave vector 
${\bf k}$ an orthonormal frame with axes dependant on ${\bf k}$ as
\begin{equation}
{\bf i}({\bf k}) = \frac{{\bf k} \times{\bf \alpha}}{|{\bf k} \times 
{\bf \alpha}|},\qquad{\bf j}({\bf k}) = \frac{{\bf k} 
    \times{\bf i}}{|{\bf k}|}, 
\qquad\frac{{\bf k}}{k};
\end{equation}
${\bf \alpha}$ is an arbitrary but fixed unit vector which we choose 
to be parallel to the rotation axis, ${\bf \alpha} = (0,0,1)$. This 
representation has been widely used to study turbulence (see, e.g., 
\cite{Lee1979,Cambon1985}). Random initial conditions are created through 
a superposition of harmonic modes with random phases $\phi({\bf k}), 
\phi'({\bf k})$ projected over the Craya reference system for each 
${\bf k}$,
\begin{equation}
{\bf u}({\bf k})=e^{i\phi}{\bf i}+e^{i\phi'}{\bf j}.
\end{equation}
Note that due to the geometry of the Craya-Herring decomposition there 
is no projection of ${\bf u}$ into ${\bf k}$ as the incompressibility 
condition stands for ${\bf u}({\bf k}) \perp {\bf k}$. 

As we are interested in the decay laws followed by flows with a particular 
initial spectrum (either Batchelor's or Saffman's), we can further 
control the shape of the initial isotropic spectrum by multiplying 
each mode ${\bf u}({\bf k})$ in Fourier space with $0 < k \le k_0$ by 
an amplitude $C k^\beta$, where C is a constant to get initial 
r.m.s.~velocity $U\approx 1$, and $\beta$ is a parameter used to adjust the 
power law in $E(k)$. For $k > k_0$, the spectrum is continued by an 
exponential decay up to $k_c$ (the maximum wavenumber explicitly resolved 
in the simulations).

\subsection{Spectra}

In order to characterize the simulations we consider isotropic as well as 
anisotropic spectra. The latter is useful to analyze rotating flows in 
which anisotropies grow with time.

In the simulations, the isotropic spectrum is defined by averaging in 
Fourier space over spherical shells,
\begin{equation}
E(k,t) = \frac{1}{2} \sum_{k\le |{\bf k}|<k+1} {\bf u}^*({\bf k},t)
\cdot {\bf u}({\bf k},t) ,
\label{eq:isoenespec}
\end{equation}
where the star denotes complex conjugate. For the anisotropic spectrum 
the components of ${\bf u}({\bf k})$ are integrated around the axis of 
rotation to obtain a spectrum that depends only on $k_{\perp}$,
\begin{equation}
E(k_{\perp}) = \frac{1}{2} \sum_{k_{\perp} \le |{\bf k_{\perp}}|<k_{\perp}+1}
{\bf u}^*({\bf k},t)\cdot {\bf u}({\bf k},t) .
\label{eq:anienespec}
\end{equation}
From now on we will reference these two spectra simply as $E(k)$ and 
$E(k_{\perp})$ respectively. 

More detailed information on the energy spectral 
distribution of anisotropic flows can be obtained from the 
axisymmetric energy spectrum $e(k_{\parallel}, k_{\perp})$ 
(see, e.g., \cite{Cambon1985,Cambon1997,Mininni2012}). 
This spectrum is obtained after integrating the three-dimensional 
energy spectrum around the axis of rotation to obtain a spectrum 
that depends only on $k_{\parallel}$ and $k_{\perp}$, and that relates to 
$E(k_{\perp})$ as follows:
\begin{equation}
E(k_{\perp})=\sum_{k_{\parallel}}{e(k_\parallel,k_\perp)}.
\end{equation}

Finally, the spectrum of the initial conditions is such that the Saffman 
or Batchelor power laws $\sim k^2$ or $\sim k^4$ are imposed in 
$E(k)$. The relationship between the power law in $E(k)$ and of 
that of $E(k_\perp)$ can be derived as follows. Equation 
(\ref{eq:isoenespec}) in the continuous limit can be expressed as
\begin{equation}
E(k,t) = \frac{1}{2}\int \hat{\Phi}_{ii}(k,t)k^2d\varOmega_k,
\label{ec:Ephi}
\end{equation}
where $\hat{\Phi}_{ij}$ is the spectral tensor, i.e., the Fourier 
transform of the second order velocity correlation function 
$\Phi_{ij} \left<u_i({\bf x},t)u_j({\bf x}+{\bf r},t)\right>$, and where 
$\varOmega_k$ is the solid angle in Fourier space. Per virtue of 
Eq.~(\ref{ec:Ephi}), if $E(k,0) \sim k^{\sigma}$ then $\hat{\Phi}_{ii} 
\sim k^{\sigma-2}$. For axisymmetric flows we can write the energy 
spectrum $E(k_{\perp})$ integrating in complex space over cylindrical 
shells, 
\begin{equation}
E(k_{\perp},t)=\frac{1}{2}\int \hat{\Phi}_{ii}(k_{\perp},t) 
    k_{\perp}d{\phi}_k dk_z.
\end{equation}
If $E(k,0)\sim k^{\sigma}$ then,
\begin{eqnarray}
E(k_{\perp},0)=\frac{1}{2}\int k^{\sigma-2}k_{\perp}d\phi_k dk_z =
    \nonumber \\
{} \frac{1}{2}\int(k^2_{\perp}+k^2_z)^{\frac{\sigma-2}{2}}k_{\perp} d\phi_k 
    dk_z,
\label{eq:alto_paper}
\end{eqnarray}

When most of the energy is in the $k_z=0$ plane then, from 
Eq.~(\ref{eq:alto_paper}) 
\begin{equation}
E(k_{\perp},0)\sim \frac{1}{l_\parallel}\int k_{\perp}^{\sigma-1}d\phi_k.
\label{eq:energiakperp}
\end{equation}

\begin{table*}
\caption{\label{table:runs}Parameters used in the simulations: number of 
linear grid points $N$,kinematic viscosity $\nu$, rotation frequency 
$\Omega$, Reynolds number $Re$, Taylor Reynolds number $Re_{\lambda}$, 
Rossby number $Ro$, micro-Rossby number $Ro^{\omega}$, and a brief 
description of the initial energy spectrum $E(k)$: the power law followed 
at large-scales and the range of scales where this power law is satisfied. 
Values of $Re$, $Re_{\lambda}$,$Ro$, and $Ro_{\lambda}$ are given at $t=0$, 
and at $t=20$ between parentheses.}
\begin{ruledtabular}
\begin{tabular}{lcccccccc}
Run&$N$&$\nu$&$\Omega$&$Re$&$Re_{\lambda}$&$Ro$&$Ro^\omega$& Initial $E(k)$ \\
\hline
A &$256$&$1.3\times 10^{-4}$& $0$ & $1580(330)$ & $780(100)$ & 
  $\infty$($\infty$)
  & $\infty$($\infty$) & $k^{2}$ $(1 \leq k \leq 40)$\\
B &$256$&$1.3\times 10^{-4}$& $33$ & $1580(1090)$ & $780(520)$ & 
  $0.15(0.004)$ & $0.9(0.024)$
  & $k^{2}$ $(1 \leq k \leq 40)$\\
C &$256$&$1.3\times 10^{-4}$& $0$ & $1280(150)$ & $730(60)$ & 
  $\infty$($\infty$)
  & $\infty$($\infty$) & $k^{4}$ $(1 \leq k \leq 40)$\\
D &$256$&$1.3\times 10^{-4}$& $33$ & $1280(360)$ & $730(180)$ & 
  $0.18(0.002)$ & $1.0(0.015)$
  & $k^{4}$ $(1 \leq k \leq 40)$
\label{tab:paratable}
\end{tabular}
\end{ruledtabular}
\end{table*}

Even in the case in which the energy is not concentrated in the 
$k_z=0$ plane (as in the isotropic initial conditions), the dependence 
of $E(k_\perp)$ will be $\sim k_\perp^{\sigma - 1}$ as it follows directly 
from Eq.~(\ref{eq:alto_paper}). To summarize,
\begin{equation}
E(k) \sim k^{\sigma} \,\, \Rightarrow \,\, E(k_{\perp}) \sim k_{\perp}^{\sigma-1}.
\label{eq:Eisoani}
\end{equation}
It is important to note that this simple relation between 
the isotropic $E(k)$ and the anisotropic $E(k_{\perp})$ spectra 
only holds for the isotropic initial conditions. However, the 
relation will be useful to understand the power laws observed 
at large scales in the following section. As an example, when 
analyzing rotating turbulence with an initial isotropic 
spectrum $E(k)\sim k^2$, we will be facing the case 
with initial $E(k_{\perp})\sim k_{\perp}$. Note the power 
laws $E(k_{\perp})\sim k_{\perp}$ and $E(k_{\perp})\sim k_{\perp}^3$, 
resulting respectively for Saffman and Batchelor isotropic initial 
conditions, are the ones that arise in the expansion of the 
axisymmetric energy spectrum for $k_\parallel=0$ and 
$k_\perp \to 0$ in Eq.~(\ref{eq:expansion2d}).

\section{\label{sec:NumRes}Numerical Results}

We now present results stemming from four numerical simulations (see 
table \ref{tab:paratable}). All runs were initialized using the random 
(Craya-Herring) initial conditions described in the previous section, 
with an initial isotropic spectrum $E(k) \sim k^2$ between $0 < k \le 40$ 
for runs A and B, and $E(k) \sim k^4$ between $0 < k \le 40$ for runs C and D. 
Runs A and C correspond to isotropic and homogeneous turbulence (no 
rotation), while runs B and D are rotating, resulting in homogeneous but 
anisotropic flows as time evolves. The purpose of simulations $A$ and $C$ 
is to recover well known results for isotropic $\sim k^2$ and $\sim k^4$ 
turbulence, and to use them as a starting point to analyze the more 
complex rotating cases. Note the decay of $\sim k^4$ turbulence 
(corresponding to runs C and D) was studied in detail in 
\cite{Teitelbaum2010}, and is considered here to compare with the 
$\sim k^2$ case.

\begin{figure}
\includegraphics[width=8cm]{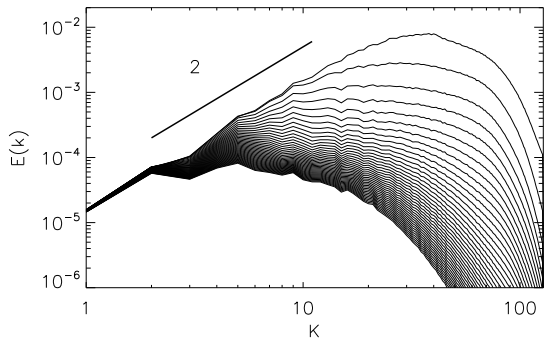}
\caption{Evolution of the isotropic energy spectrum $E(k)$ for run A from 
$t=0$ to $t=20$ with time increments $\Delta t=0.5$. Note how the initial 
$E \sim k^2$ spectrum is preserved over time at the larges scales (smaller 
wave numbers).}
\label{fig:espectrum_norot}
\end{figure}

Parameters of the runs listed in table \ref{tab:paratable} are defined 
as follows: The Reynolds number is
\begin{equation}
Re = \frac{lU}{\nu}
\end{equation}
where $U$ is the r.m.s velocity and the energy-containing 
scale $l$ is defined from the isotropic energy spectrum $E(k)$ as 
\begin{equation}
l=2\pi \frac{\int E(k)k^{-1}dk} {\int E(k)dk}. 
\end{equation}
The Reynolds number based on the Taylor scale $\lambda$ is
\begin{equation}
Re_\lambda = \frac{\lambda U}{\nu}
\end{equation}
where 
\begin{equation}
\lambda = 2 \pi \left(\frac{\int E(k)dk}{\int E(k)k^2dk}\right)^{1/2}. 
\end{equation}
The Rossby number is
\begin{equation}
Ro = \frac{U}{2 \Omega l},
\end{equation}
while the micro-Rossby number is defined as
\begin{equation}
Ro^\omega = \frac{\omega}{2 \Omega}.
\end{equation}
This number should be initially of order one for the scrambling effect of 
waves not to completely damp the non-linear term in the Navier-Stokes 
equation, which would result in just an exponential decay of 
the energy \cite{Cambon1997}.

Time-dependent parameters in table \ref{tab:paratable} are given for all 
runs at $t=0$, and also at $t=20$ (roughly ten turnover times after the 
self-similar decay of energy starts).

\begin{figure}
\includegraphics[width=8cm]{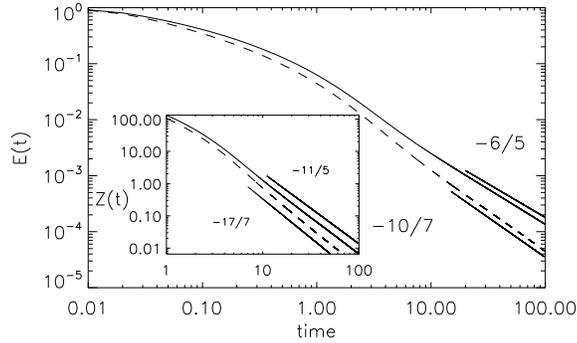}
\caption{Energy evolution for run A (solid) and C (dashed). Note the 
decay laws with exponents $\approx -6/5$ and $\approx -10/7$ respectively, 
after an initial transient of $\approx 10$ turn over times. Slopes 
are shown as a reference. Inset: enstrophy decay for the same runs. 
The decays approach $\approx -11/5$ and $\approx -17/7$ respectively after 
$t\approx 10$.}
\label{fig:decay_norot} 
\end{figure}

\subsection{Decay of non-rotating flows}

Figure \ref{fig:espectrum_norot} shows the evolution of the isotropic 
energy spectrum for run A, at times ranging from $t=0$ to $t=20$. The 
initial $\sim k^2$ law is approximately preserved over time at the 
smallest wavenumbers, while higher wavenumbers decay due to the effect 
of dissipation. This behavior is consistent with the quasi-conservation 
of $L$ in Eq. (\ref{eq:Elowk}), for which a subsequent decay law 
$E \sim t^{-6/5}$ is expected for the energy.

\begin{figure}
\includegraphics[width=8cm]{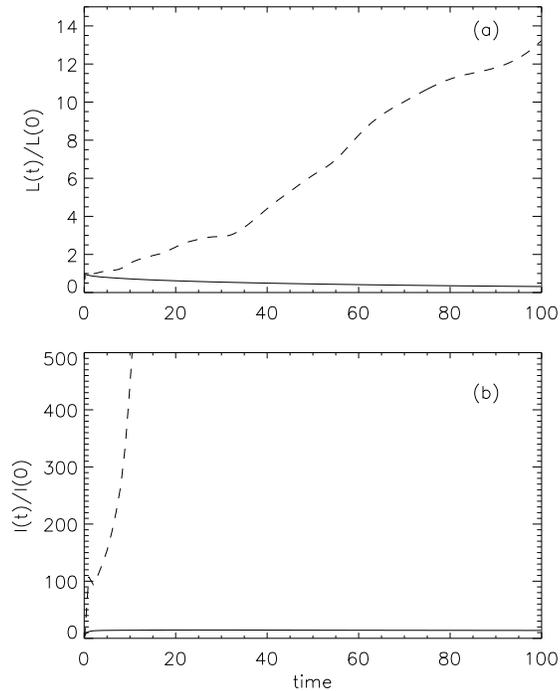}
\caption{(a) Evolution of the integral $L$ normalized by its initial value 
$L(0)$ for the isotropic run A (solid) and rotating run B (dashed).  
(b) Evolution of the integral $I$ normalized by its initial value $I(0)$ 
for the isotropic run B (solid) and rotating run D (dashed).}
\label{fig:L_vs_time}
\end{figure}

Figure \ref{fig:decay_norot} shows the energy and enstrophy decay for 
runs A and C. After a transient lasting for approximately $10$ turn-over 
times, both runs decay with different power laws. In run A, the energy 
decays close to $E \sim t^{-6/5}$ as predicted by Eq. (\ref{eq:teodecay}), 
and as expected from the shape of the large-scale spectrum in Fig. 
\ref{fig:espectrum_norot}. The enstrophy shows a faster decay close to 
$Z \sim t^{-11/5}$, consistently with phenomenological arguments which 
indicate that $E \sim t^{\alpha}$ leads to $Z \sim t^{\alpha -1}$ under 
the assumption of isotropy, as
\begin{equation}
\frac{dE}{dt} = -2 \nu Z(t) .
\label{eq:Zdecay}
\end{equation}

The energy decay for run C is consistent with $E \sim t^{-10/7}$ as 
expected by Eq. (\ref{eq:teodecay}), while the enstrophy decay is 
close to $Z \sim t^{-17/7}$, also consistent with Eq.~(\ref{eq:Zdecay}). 
The spectrum of run C maintains a $\sim k^4$ shape for large scales (not 
shown). All these results are consistent with previous simulations and 
theoretical results \cite{Kolmogorov1941,Ishida2006,POF}. 

The prediction $E \sim t^{-6/5}$ for run A assumes $L$ remains constant 
during the decay. To verify this, and to further study whether 
$L$ also remains constant in the decay of rotating turbulence with 
$\sim k^2$ spectrum, we show the time evolution of $L$ for runs A and B 
in Figure \ref{fig:L_vs_time}(a). We estimated $L$ in two different ways: 
A fit to the spectrum with a power law $\sim k^2$ was computed for the smallest 
wavenumbers to obtain the multiplicative prefactor in Eq.~(\ref{eq:Elowk}) 
(proportional to $L$ if the spectrum is $\sim k^2$), as was done 
in \cite{Ishida2006}. We also computed the two-point correlation function 
from the energy spectrum using \cite{Davidson2004} 
\begin{equation}
\left < {\bf u \cdot u'} \right > (r)=2\int_0^{\infty} 
E(k)\frac{\sin (kr)-kr \cos (kr)}{(kr)^3}dk,
\end{equation}
and then used Eq.~(\ref{eq:L}) to compute $L$. Both methods give consistent 
results and in the following we show results obtained with the last method.

\begin{figure}
\includegraphics[width=8cm]{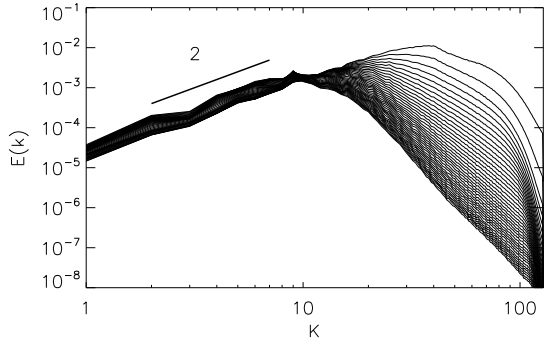}
\caption{Evolution of the isotropic energy spectrum $E(k)$ for run B 
from $t=0$ to $t=20$ with time increments $\Delta t=0.5$. Note how the 
initial $E \sim k^2$ spectrum for large scales (low wave numbers) is 
preserved over time, but its overall amplitude increases.}
\label{fig:spectrum_rot_iso} 
\end{figure}

In run A, $L$ decays to half its initial value between $t=0$ and 
$t\approx 10$, but after this time it only changes slowly with time and 
can be considered almost constant (note the energy between $t=10$ 
and $t=100$ changes by more than an order of magnitude). However, $L$ is far 
from being constant in run B; the behavior of the rotating runs is 
discussed in more detail in the next subsection.

Similar results were found for the evolution of $I$ in simulations C 
and D, as shown in Fig.~\ref{fig:L_vs_time}(b). In run C, $I$ grows during 
a short transient, but stays constant during the self-similar decay of 
the energy.

While the behavior of $L$ and $I$ in the simulations is compatible with 
the theoretical and phenomenological arguments discussed in 
Sec.~\ref{sec:Karman}, the rapid growth of these quantities in the 
runs with rotation seems to question the use of isotropic integrals 
to derive decay laws for anisotropic flows.

\begin{figure}
\includegraphics[width=8cm]{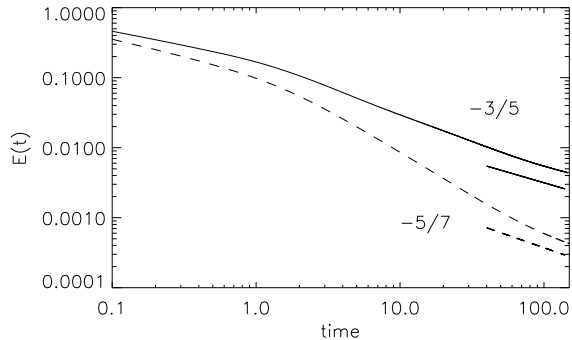}
\caption{Energy decay for runs B (solid) and D (dashed); $t^{-3/5}$ 
and $t^{-5/7}$ power laws are shown as a reference, corresponding to 
phenomenological predictions for Saffman and Batchelor rotating 
turbulence based on isotropic conserved quantities.}
\label{fig:Exy_rot_kz=0} 
\end{figure}

\subsection{\label{sec:RotFlow}Decay of rotating flows}

The analysis for runs A and C was based on the isotropic energy 
spectrum and on the quasi-invariance of the isotropic integrals 
$L$ and $I$. However, the growth of $L$ ($I$) in run B (D) suggests 
that different arguments should be used to predict their energy 
decay. In this light, we now consider the anisotropic quantities 
introduced in Sec.~\ref{sec:Karman}, $L_{\perp}$ and $I_{\perp}$, and 
show that further analysis can be done by studying the evolution of 
anisotropic spectra and splitting the energy decay in $2D$ and $3D$ 
modes. A similar method was used in \cite{POF} to study $\sim k^4$ 
turbulence.

The evolution of the isotropic spectrum for run B is shown in 
Fig.~\ref{fig:spectrum_rot_iso}, for different times from $t=0$ to 
$t=20$. As for run A, its initial $\sim k^2$ behavior for large scales 
is preserved over time. However, its amplitude at large scales 
increases with time, in agreement with the evolution of $L$.

\begin{figure}
\includegraphics[width=8cm]{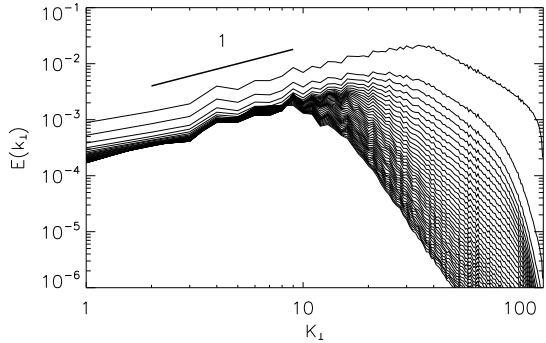}
\caption{Evolution of the axisymmetric energy spectrum $E(k_{\perp})$ 
for run B from $t=0$ to $t=20$ with time increments $\Delta t=0.5$. 
The initial shape $E \sim k_{\perp}$ of spectrum at large scales (low 
wavenumbers) is approximately preserved over time.}
\label{fig:spectrum_rot_ani}
\end{figure}

The time evolution of the total energy in run B shows that 
the decay is close to, but steeper than, $E \sim t^{-3/5}$. In Figure 
\ref{fig:Exy_rot_kz=0} we show the energy decay for this run (we also 
show the $\sim k^4$ case for comparison, i.e., run D). The decay deviates 
from the expected $\sim t^{-3/5}$ law derived via isotropic arguments (in 
the same fashion, run D deviates from the expected $\sim t^{-5/7}$ law) 
and strictly speaking, a power law is hard to identify. Indeed, as was 
shown in Fig.~\ref{fig:L_vs_time}, both $L$ and $I$ grow fast in runs 
B and D. As a result, arguments based on the quasi-conservation of 
$L$ or $I$ should be expected to fail to predict the correct energy 
decay. The reason for this failure can be associated with the fact that 
rotating flows are not isotropic, and instead become axisymmetric with 
a tendency towards two-dimensionalization.

From the arguments in Sec.~\ref{sec:Karman}, we expect only large-scale 
correlations in the direction perpendicular to the axis of rotation 
to be preserved, and either $L_\perp$ or $I_\perp$ to remain constant 
depending on the initial conditions. Therefore, we analyze runs B and D 
with the aid of the expansion in Eq.~(\ref{eq:expansion2d}) and of the 
anisotropic energy spectrum $E(k_{\perp})$. Following Eqs.~(\ref{eq:E2D}) 
and (\ref{eq:teodecayrot}), we will also separate the energy into the 
energy of slow modes $E_{2D}$, and the energy of fast modes $E_{3D}$.

\begin{figure}.
\includegraphics[width=8cm]{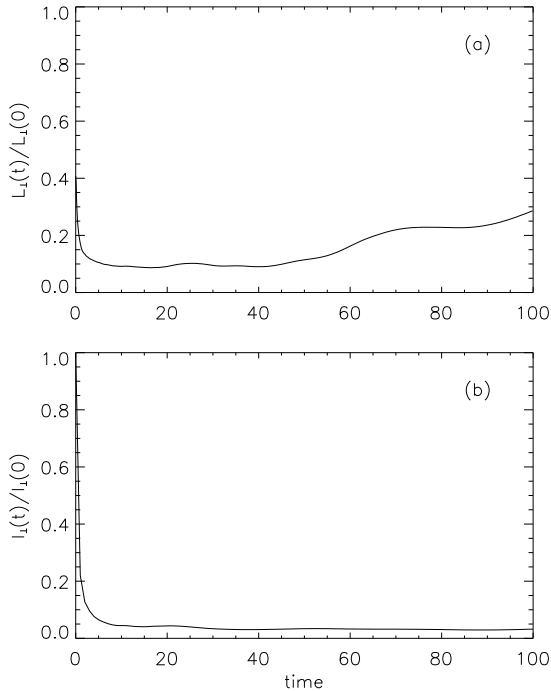}
\caption{(a) Ratio $L_{\perp}(t)/L_{\perp}(0)$ as a function of time 
for run B. (b) Evolution of $I_{\perp}(t)/I_{\perp}(0)$ in run D. Note 
that in both cases, these magnitudes remain approximately constant after 
a short initial transient.}
\label{fig:K_vs_time} 
\end{figure}

In Fig.~\ref{fig:spectrum_rot_ani} we show the time evolution of 
$E(k_{\perp})$ in run B. Per virtue of Eq.~(\ref{eq:Eisoani}), as the 
isotropic spectrum is $E(k)\sim k^2$ for small wavenumbers, the 
anisotropic spectrum is $E(k_{\perp})\sim k_{\perp}$. As for $E(k)$, 
the shape of the spectrum is preserved during the decay, and 
$E(k_{\perp})\sim k_{\perp}$ for small $k_{\perp}$ at all times. However, 
unlike the isotropic spectrum, $E(k_{\perp})$ rapidly decreases its 
amplitude during a short transient, and then the amplitude of the 
spectrum at large scales stabilizes and varies only slowly with time. 
Run D shows a similar behavior in $E(k)$ and $E(k_{\perp})$, but 
following $E(k_{\perp})\sim k_{\perp}^3$ instead.

We saw in Sec.~\ref{sec:Karman} that constancy of $L_{\perp}$ ($I_{\perp}$) 
may be expected for $E(k) \sim k^2$ ($\sim k^4$) initial spectra if 
large-scale correlations decay fast enough, and that phenomenological 
analysis leads then to $E_{2D}(t) \sim t^{-1/2}$ ($E_{2D} \sim t^{-2/3}$). 
As for $L$, we estimate $L_{\perp}$ using two different methods: 
By fitting the spectrum for small $k_\perp$, and by using Eq.~(\ref{eq:K}). 
The two-point longitudinal correlation function for the axisymmetric 
case is estimated using Bessel functions and the anisotropic perpendicular 
spectrum (see, e.g., \cite{Davidson2004,Teitelbaum2010}),
\begin{equation}
\left < {\bf u_{\perp} \cdot u_{\perp}'} \right> 
(r_{\perp})=2\int E(k_{\perp})J_0(k_{\perp}r_{\perp})dk_{\perp}.
\label{eq:corrperp}
\end{equation}
Both estimations give similar results and the curves discussed below 
are obtained from Eqs.~(\ref{eq:K}) and (\ref{eq:corrperp}). The same 
procedure was used to estimate $I_\perp$  from Eq.~(\ref{eq:I2D}).

In Fig.~\ref{fig:K_vs_time} we show the evolution of $L_{\perp}$ and 
$I_{\perp}$ normalized by their values at $t=0$, for runs B and D 
respectively. Although at early times their values decrease rapidly, 
they remain afterwards approximately constant during the entire 
simulation (note that in the same simulations and at the same times, 
$L$ and $I$ increase by at least one order of magnitude).

\begin{figure}
\includegraphics[width=8cm]{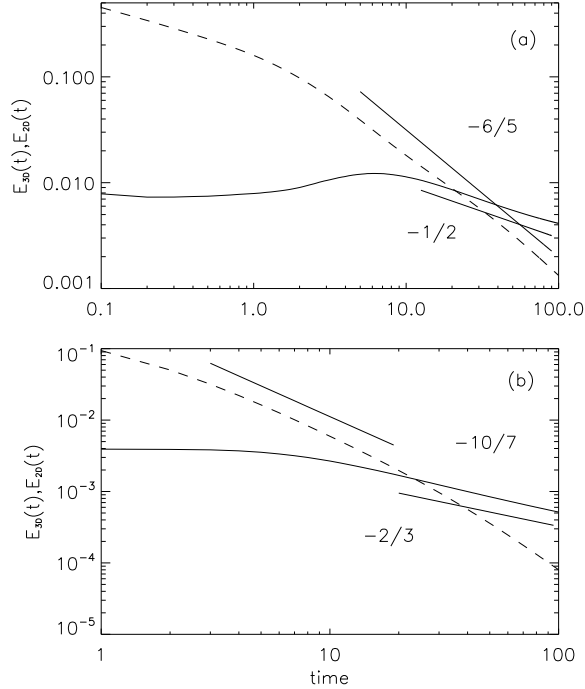}
\caption{(a) Time evolution of the energy $E_{3D}$ in modes with 
$k_{\parallel} \neq 0$ (dashed), and of the energy $E_{2D}$ in modes with 
$k_{\parallel}=0$, for run B. Power laws $t^{-6/5}$ and $t^{-1/2}$ 
are indicated as a reference, following phenomenological arguments 
respectively for the decay of the 3D energy, and for the decay of the 2D 
energy based on approximate constancy of $L_\perp$ in Saffman turbulence. 
(b) Same for run D. The power laws $t^{-10/7}$ and $t^{-2/3}$ correspond 
to the phenomenological predictions for Batchelor turbulence.}
\label{fig:energias2d3d} 
\end{figure}

The evolution of the energy for run B shows interesting properties (see 
Fig.~\ref{fig:energias2d3d}); $E_{2D}$ initially grows until 
$t \approx 10$, when it reaches its maximum value and begins to decrease. 
$E_{3D}$ decreases faster than $E_{2D}$ and at $t\approx 20$ both energies 
are comparable. After that time, the system is dominated by the energy 
in the slow modes. Starting at $t \approx 10$, both energies show a 
decay compatible with power laws with different exponents. $E_{2D}(t)$ 
decays close to $\sim t^{-1/2}$, in agreement with 
Eq.~(\ref{eq:teodecayrot}) and with the approximate constancy of 
$L_\perp$,  while $E_{3D}$ decays close to $\sim t^{-6/5}$, the value 
expected from Eq.~(\ref{eq:teodecay}) for the decay of three-dimensional 
Saffman turbulence.

A distinct decay of $E_{2D}$ and $E_{3D}$ is also observed in run D 
(corresponding to $\sim k^4$ turbulence, also shown in 
Fig.~\ref{fig:energias2d3d}). The decay of the energy in slow modes 
is compatible with $E_{2D}(t) \sim t^{-2/3}$, as expected for 
Batchelor turbulence from Eq.~(\ref{eq:teodecayrot}) and with the 
approximate constancy of $I_\perp$ shown in Fig.~\ref{fig:K_vs_time}. 
The 3D energy shows a decay that is close to $E_{3D} \sim t^{-10/7}$.

\begin{figure}
\includegraphics[width=8cm]{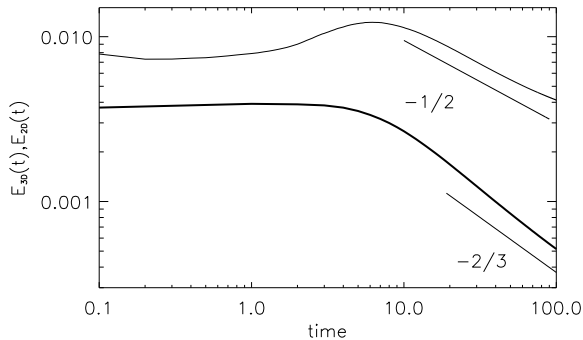}
\caption{$E_{2D}$ decay as a function of time for runs B (thin) and D 
(thick). The expected decays based on constancy of $L_\perp$ and $I_\perp$ 
are indicated as a reference. The agreement in this case is much better 
than when assuming constancy of the isotropic integrals $L$ and $I$ (see 
Fig.~\ref{fig:Exy_rot_kz=0}).}
\label{fig:2Ddecays}
\end{figure}

The decay of $E_{2D}$ in runs B and D is compared in 
Fig.~\ref{fig:2Ddecays}. The two different decay laws followed by the 
energy in the two runs can be clearly identified. The result indicates 
that the decay of rotating turbulence is affected by large scale 
correlations, as different power laws can be observed for 
$E(k) \sim k^2$ and $\sim k^4$ initial spectra.

The different power laws followed by $E_{2D}$ and $E_{3D}$ after 
$t \approx 10$ in both runs point to a negligible interchange of energy 
between slow and fast modes at late times, as required for an equation 
like Eq.~(\ref{eq:E2D}) to hold. Runs B and D have initial values of 
the Rossby number which correspond to the intermediate Rossby range 
studied in \cite{Bourouiba2007}. In this range, maximal energy transfer 
from 3D to 2D modes takes place at early times, with $E_{2D}$ in some 
cases growing from energy in 3D modes as a result (see 
Fig.~\ref{fig:energias2d3d} (a)). After this phase, the independent 
decay of the two energies implies that these exchanges are small 
compared with, e.g., the transfer of energy from vortical motions at 
large scales to vortical motions at smaller scales in $E_{2D}$, and 
the transfer from 3D modes to 3D modes in $E_{3D}$.

\begin{figure}
\includegraphics[width=8cm]{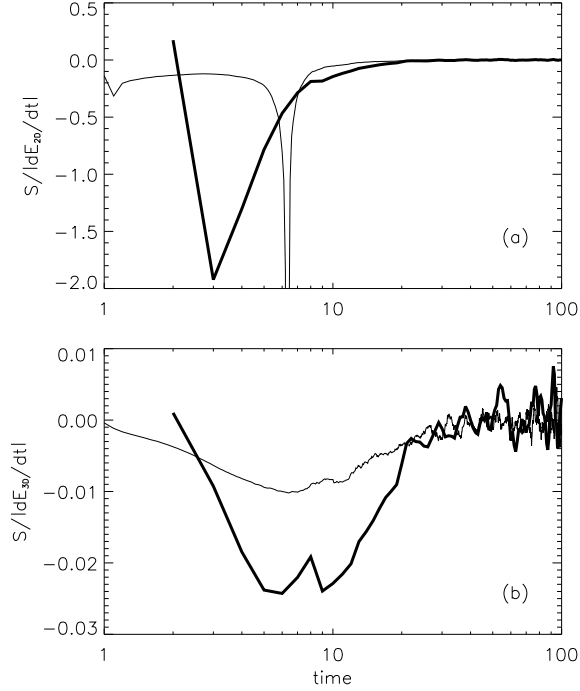}
\caption{(a) Energy interchanged per unit of time $S$ between 2D and 3D 
modes, (a) normalized by $|dE_{2D}/dt|$, and (b) normalized by $|dE_{3D}/dt|$, 
for runs B (thin) and D (thick). Positive values of S represent flux 
of energy from 2D to 3D modes, while negative values indicate flux 
from 3D to 2D motions. At early times $S$ is negative, indicating 
energy is transferred from 3D to 2D modes. After $t \approx 10$ in run 
B, and $t \approx 10$ in run D, when the distinct power law decays are 
observed for $E_{2D}(t)$ and $E_{3D}(t)$ (Fig.~\ref{fig:energias2d3d}), 
$S$ becomes negligible.}
\label{fig:S}
\end{figure}

\subsection{Energy exchange between 2D and 3D modes}

A measure of the exchange of energy between 2D and 3D modes per unit 
of time can be obtained from the flux of energy across planes in Fourier 
space with normal $k_z$. The amount of energy per unit of time going 
through any of such planes is given by
\begin{equation}
\Pi(k_\parallel) = - \iiint_{k_\parallel=0}^{k_\parallel} \hat{\bf u}^*_{\bf k} \cdot 
    \widehat{\left({\bf u} \cdot \nabla {\bf u}\right)}_{\bf k} \, 
    dk_x dk_y dk_z ,
\end{equation}
where the integrals in $k_x$ and $k_y$ run over their entire range 
(see \cite{Mininni2011} for definitions of anisotropic fluxes and 
spectral transfer functions), and the hat denotes Fourier transform. 
The energy transferred from 2D modes to 3D modes per unit of time is 
then
\begin{equation}
S = \Pi(k_\parallel=0) .
\end{equation}
When S is positive, it represents transfer of energy from 2D to 3D 
motions (i.e., energy going from $k_\parallel=0$ towards larger values of 
$k_\parallel$, resulting in an effective source of 3D energy coming from 
2D loses), while when negative it represents transfer from 3D to 2D 
motions (i.e., a source of 2D energy from 3D loses). For these sources 
(or loses) of energy to be negligible, they must be small when compared 
with the time derivatives $dE_{2D}/dt$ and $dE_{3D}/dt$. When they are, 
the decay of $E_{2D}$ and of $E_{3D}$ can be considered independently.

\begin{figure}
\includegraphics[width=8.5cm]{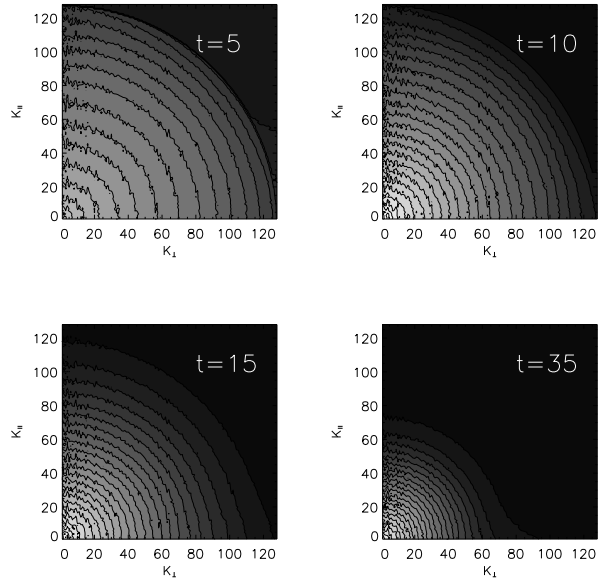}
\caption{Axisymmetric energy spectrum $e(k_{\parallel}, k_{\perp})/ \sin \theta$ 
at different times in run A. The circular contour levels indicate an isotropic 
energy distribution.}
\label{fig:contourA}
\end{figure}

\begin{figure}
\includegraphics[width=8.5cm]{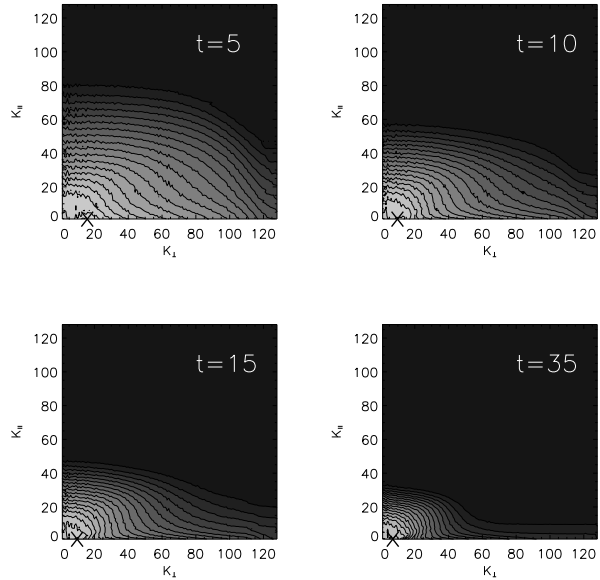}
\caption{Axisymmetric energy spectrum $e(k_{\parallel}, k_{\perp})/ \sin \theta$ 
at different times in run B. For rotating flows, the spectral
distribution of energy becomes anisotropic with more energy near 
the $k_{\parallel}= 0$ plane. The cross indicates the maximum of the 
spectrum at the different times.}
\label{fig:contourB}
\end{figure}

Figure \ref{fig:S} shows the function $S$ normalized by $|dE_{2D}/dt|$ 
and $|dE_{3D}/dt|$ for runs B and D. In both runs $S$ is negative 
before $t\approx 10$, indicating energy goes from 3D to 2D modes, 
and explaining the growth of $E_{2D}$ at early times in run B. Note that 
in run B, $|dE_{2D}/dt|$ becomes zero at $t\approx 5$, resulting in an 
infinite value of the ratio $S/|dE_{2D}/dt|$ in Fig.~\ref{fig:S} (a). Later, 
for $t \approx 10$ in run B, and for $t \approx 20$ in run C, 
$S/|dE_{2D}/dt|$ and $S/|dE_{3D}/dt|$ become small, indicating the 
interchange of energy between slow and fast modes becomes negligible 
when compared with the energy decay rates, and justifying the use 
of separate balance equations to study the decay of $E_{2D}$ and $E_{3D}$. 
Indeed, to have $S \ll |dE_{2D}/dt|$ and $S \ll |dE_{3D}/dt|$ is enough 
to consider equations for the evolution of $E_{2D}$ and $E_{3D}$ as was 
done in Sec.~\ref{sec:Karman}, as is a more precise definition of 
what was meant in that section by the condition of ``weakly coupled'' 
2D and 3D modes.

It is worth pointing out that the fact that $S/|dE_{2D}/dt|$ and 
$S/|dE_{3D}/dt|$ are much smaller than unity does not imply that the 2D and 
3D modes are completely decoupled, and only implies that almost no energy 
is interchanged between slow and fast modes during the self-similar decay. 
Triadic interactions, e.g., between two fast modes and one slow mode, can 
still occur and be relevant, as long as the slow mode is only an intermediary 
and does not receive from nor give net energy to the fast modes. Indeed, such 
interactions between slow and fast modes that preserve the 2D and 3D energy 
separatedly would be required to have $dE_{2D}/dt$ proportional to 
$1/\Omega$ in Eq.~(\ref{eq:balancerot2D}), as in their absence the decay 
of $E_{2D}$ should be that of 2D Navier-Stokes and independent of 
$\Omega$ (note triadic interactions between three slow modes are 
trivially resonant and independent of $\Omega$).

\begin{figure}
\includegraphics[width=8.5cm]{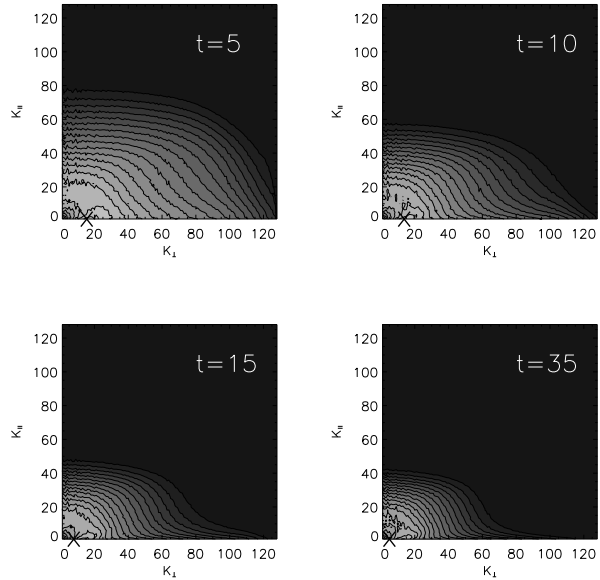}
\caption{Axisymmetric energy spectrum $e(k_{\parallel}, k_{\perp})/ \sin \theta$ 
at different times in run D. As in run B, the spectral energy
distribution is anisotropic.  The cross indicates the maximum of the 
spectrum at the different times.}
\label{fig:contourD}
\end{figure}

\subsection{Energy spectral distribution}

A deeper understanding of the development of anisotropy in 
the flow cannot be obtained solely from studying the reduced 
spectrum $E(k_\perp)$ and the isotropic spectrum $E(k)$. To 
further investigate the energy spectral distribution, we present 
contour plots of the axisymmetric energy spectrum 
$e(k_{\parallel}, k_{\perp})$ for runs A, B, and D in 
Figs.~\ref{fig:contourA}, \ref{fig:contourB}, and \ref{fig:contourD} 
respectively. Note that in order to obtain circular contour levels
when the spectral distribution of the energy is isotropic, in these 
figures the axisymmetric energy spectrum is divided by 
$\sin \theta$, where $\theta = \arctan(k_{\parallel}/k_{\perp})$.

\begin{figure}
\includegraphics[width=8cm]{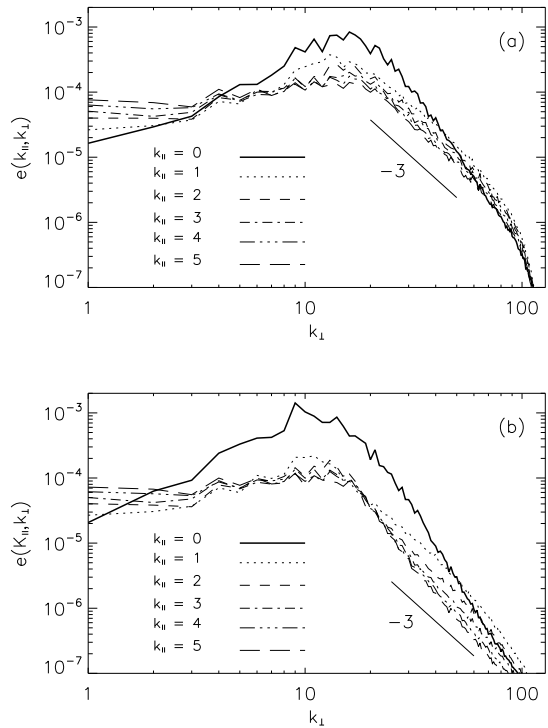}
\caption{Axysimmetric energy spectrum $e(k_{\parallel}, k_{\perp})$ for 
run B for different values of $k_{\parallel}$ at (a) $t=5$, and (b) $t=10$. A $k_{\perp}^{-3}$ slope is shown as a reference.}
\label{fig:kparaB}
\end{figure}

In the absence of rotation (Fig.~\ref{fig:contourA}) the spectrum 
shows an isotropic energy distribution evidenced by circular contour levels 
which maintain their shape as the flow decays. However, when rotation 
is present, the spectral distribution of the energy becomes
anisotropic with more energy near the $k_{\parallel}= 0$ axis (see 
Figs.~\ref{fig:contourB} and \ref{fig:contourD}). Already at $t=10$, the 
maximum of the axisymmetric energy spectrum takes place in 
the $k_{\parallel}= 0$ axis, in agreement with the preferential
transfer from 3D modes towards slow 2D modes observed at early 
times in the flux in Fig.~\ref{fig:S} (as a matter of fact, the
maximum of the spectrum is already located in the $k_{\parallel}= 0$ 
axis for times as early as $t=5$).

After $t=10$, the exchange of energy between 2D and 3D modes 
becomes negligible. As already observed in the time evolution of
$E_{2D}$ and $E_{3D}$, the energy in the 3D modes decays faster than
the energy in 2D modes. This results in an increase of the spectral 
anisotropy as time evolves, with most of the energy near the 
$k_{\parallel}= 0$ axis at late times. At the same time, the peak of
the spectrum slowly moves through the $k_{\parallel}= 0$ axis towards 
smaller values of $k_\perp$, although even at late times the peak is 
sufficiently far from $k_\perp=1$.

\begin{figure}
\includegraphics[width=8cm]{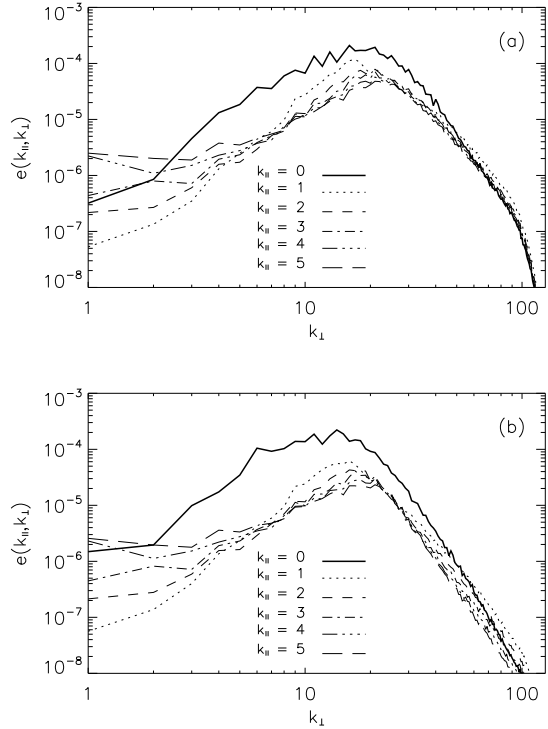}
\caption{Axysimmetric energy spectra $e(k_{\parallel}, k_{\perp})$ for 
run D for different values of $k_{\parallel}$, at (a) $t=5$, and (b)
$t=10$.}
\label{fig:kparaD}
\end{figure}

The shape of the spectrum near the $k_\parallel=0$ plane is of 
interest for many theories of homogeneous rotating turbulence (see, 
e.g., \cite{Cambon2004,Bellet2006}). It is important to note that here 
the domain with finite size results in a discrete set of values for
$k_\parallel$ so the $k_\parallel \to 0$ limit cannot be studied,
unlike homogeneous flows in infinite domains for which 
$k_\parallel$ is a continuum. Bearing in mind this limitation, we show 
in Figs.~\ref{fig:kparaB} and \ref{fig:kparaD} the axisymmetric 
energy spectrum $e(k_{\parallel}, k_{\perp})$ for fixed values of 
$k_{\parallel}$ (from 0 to 5), for runs B and D at two different times
($t=5$ and $t=10$).

Figures \ref{fig:kparaB} and \ref{fig:kparaD} confirm that 
most of the energy is contained in modes with $k_{\parallel}=0$ 
(2D modes), and that the wavenumber $k_\perp$ at which the peak 
of the spectrum takes place slowly decreases with time (although 
the wavenumber is larger than $k_\perp=1$). As an example, in run 
B the peak of the spectrum for $k_{\parallel}=0$ is at 
$k_{\perp} \approx 20$ at $t = 5$, at $k_{\perp} \approx 15$ at 
$t=10$, and at $k_{\perp} \approx 8$ at $t = 35$. The spectrum 
for modes with $k_{\parallel}=0$ also shows a positive slope for 
wavenumbers smaller than the energy containing wavenumber, 
and the slope seems to be approximately preserved as the system 
evolves in time.

The spectra $e(k_{\parallel}, k_{\perp})$ for 
$k_{\parallel} \geq 1$ are approximately flat for small wavenumbers (specially for 
run B), and show a slope compatible with a $k_{\perp}^{-3}$ power law 
(specially for early times, at $t=5$, and for the spectrum 
corresponding to $k_{\parallel} = 1$). This is consistent with the 
scaling $e(k_{\parallel}, k_{\perp}) \sim k_{\parallel}^{-1/2}k_{\perp}^{-3}$ 
valid for $k_\parallel/k_\perp \ll 1$ 
derived from inertial wave turbulence equations in 
\cite{Cambon2004}, and also observed before in numerical 
simulations in \cite{Thiele2009}.

\section{\label{sec:conclusions}Conclusions}
We have analyzed numerical simulations of freely decaying rotating 
turbulence in periodic domains in the regime of intermediate Rossby number 
($Ro \approx 0.1$), with initial energy spectra $\sim k^2$ and $\sim k^4$. 
We first presented a brief theoretical discussion of conserved integral 
quantities for rotating turbulence using a von K\'arm\'an-Howarth 
equation which includes the Coriolis term due to rotation, extending 
previously derived results for Saffman turbulence \cite{Davidson2010}, 
considering also the Batchelor spectrum. Assuming 2D and 3D modes are 
only weakly coupled in rotating turbulence, constancy of the anisotropic 
integral quantities $L_{\perp}$ or $I_{\perp}$ was used to derive 
phenomenologically the energy decay rate expected for energy in 2D 
modes.

In simulations without rotation, we recovered well known results for 
isotropic $\sim k^2$ and $\sim k^4$ turbulence, and also verified the 
approximate constancy of the isotropic integrals $L$ and $I$. 
On the other hand, in the simulations with rotation a self-similar 
decay was difficult to identify for the total energy, while $L$ and $I$ 
were found to grow rapidly during the entire decay. However, the decay of 
energy in two-dimensional and three-dimensional modes followed distinct 
power laws, in agreement with the phenomenological predictions and with 
constancy of $L_\perp$ and $I_\perp$, which were also observed to remain 
approximately constant in the simulations.

The separate power laws followed by $E_{2D}(t)$ and $E_{3D}(t)$ in 
decaying rotating turbulence require the interchange of energy between 
slow and fast modes to be negligible. We verified that this was the case by 
studying the flux of energy across planes with normal $k_z$ in Fourier space. 
The results showed that energy is initially transferred from 3D to 2D 
modes, as expected for rotating turbulence at intermediate Rossby number 
\cite{Bourouiba2007}, but that later the flux of energy between 3D and 
2D modes becomes negligible. Contour levels of the axisymmetric 
energy spectral distribution show that in presence of rotation the
energy accumulates rapidly near the $k_{\parallel}=0$ plane, and that
at later times the anisotropy increases as a result of the different
decay rates of the $E_{2D}$ and $E_{3D}$ components of the total
energy.

Altogether, the results show that rotating turbulence is indeed affected 
by initial large scale correlations in the flow, as different decay laws 
arise for $\sim k^2$ and $\sim k^4$ initial spectra.

\begin{acknowledgments}
The authors acknowledge support from UBACYT Grant No.~20020090200692, PICT 
Grant No.~2007-02211, and PIP Grant No.~11220090100825.
\end{acknowledgments}


\end{document}